# Ein neues Kodierschema als Basis für die klinische Beurteilung frühkindlicher Vokalisationen und Deep Acoustic Phenotyping

Tomas Kulvicius[1,2*], Sigrun Lang[1*], Claudius AA Widmann[1], Nina Hansmann[1], Daniel Holzinger[3], Luise Poustka[1,4], Dajie Zhang[1,4,5#], Peter B Marschik[1,4,5,6#]

[1]Klinik für Kinder- und Jugendpsychiatrie und Psychotherapie, Universitätsmedizin Göttingen, Deutschland

[2]Abteilung für Computational Neuroscience, Drittes Physikalisches Institut, Georg-August-Universität Göttingen, Deutschland

[3]Forschungsinstitut für Entwicklungsmedizin, Johannes Kepler Universität, Linz, Österreich

[4]Leibniz ScienceCampus Primate Cognition, Göttingen, Deutschland

[5]iDN – interdisciplinary Developmental Neuroscience, Klinische Abteilung für Phoniatrie, Medizinische Universität Graz, Österreich

[6]Center of Neurodevelopmental Disorders (KIND), Karolinska Institutet, Stockholm, Schweden

*ErstautorInnen

#LetztautorInnen

**Zusammenfassung:** *Theoretischer Hintergrund*: Die frühe verbale Entwicklung ist vor allem in ihrer Entstehungsphase noch nicht vollständig verstanden. *Fragestellung*: lässt sich ein zuverlässiges, einfach anzuwendendes Kodierschema zur Klassifizierung frühkindlicher Vokalisationen definieren, das als Grundlage für die weitere Analyse der Sprachentwicklung anwendbar ist? *Methode*: In einer Längsschnittstudie mit 45 neurotypischen Säuglingen analysierten wir Vokalisationen der ersten 4 Lebensmonate. Audiosegmente wurden 5 Klassen zugeordnet: (1) Stimmhafte und (2) Stimmlose Lautäußerung; (3) Definiertes Signal; (4) Non-Target; (5) Nicht zuordenbar. *Ergebnisse*: Zwei Kodiererinnen mit unterschiedlicher Erfahrung erzielten ohne intensives Training hohe Übereinstimmung. *Diskussion und Schlussfolgerung*: Das reliable Schema kann in Forschung und Klinik für die effiziente Kodierung kindlicher Vokalisationen eingesetzt werden, als Grundlage für detaillierte manuelle und maschinelle Analysen.

**Schlüsselwörter:** Frühkindlich, Sprachentwicklung, Vokalisation, Babbeln, Kodierschema

# Facilitating deep acoustic phenotyping: A basic coding scheme of infant vocalisations preluding computational analysis, machine learning and clinical reasoning

**Abstract:** *Theoretical Background*: Very early infant development is still underresearched and emerging speech-language functions are underexplored. A reliable easy-to-apply annotation tool for very early infant vocalisations is lacking. *Objective*: To provide a tool, a ground-annotation scheme/coding scheme, to classify infant vocal data as basis for in-depth analysis of emerging verbal functions.

*Method*: Longitudinal data from a prospectively recruited neurotypical infant cohort (N = 45) was analysed. Spontaneous movements and vocalisations were fortnightly filmed in seven consecutive lab sessions from the $4^{th}$-$16^{th}$ week post-term age. Audio sequences were then semi-automatically segmented. A total of 9305 audio segments including both infant vocalisations and other sounds were exported for human annotation. A coding scheme with 5 classes was defined: (1) voiced vocalisation; (2) unvoiced vocalisation; (3) fixed signal; (4) non-target: sound not vocalised by the infant; and (5) infant vocalisation that cannot be assigned with certainty to classes (1)-(3). Two coders, one with rich experience in coding vocalisations, the other with no experience, were involved. The coders were instructed and familiarised with the scheme. No intensive training was needed for the coders to achieve high agreements on the pilot datasets. The coders went on to annotate the 9305 segments. The consensually annotated vocalisations were computationally analysed to exemplify the use of the dataset generated with the proposed coding scheme. *Results*: For the 9305 segments annotated by both coders, the 5-class Cohen's Kappa was .70, and .92 for classes (1)-(4). The test-retest reliability of the experienced coder was .82, and .68 of the novice coder. Of the 9305 segments, 27.6% was consensually classified as voiced vocalisations, 0.1% as unvoiced vocalisations, and 1.5% as fixed signals. The computational analysis replicated findings that the fundamental frequency and the duration of vocalisations are comparable between baby boys and girls. *Discussion and Conclusion*: The proposed coding scheme has proved applicable for both the experienced and the novice coder from-scratch. In contrast to other schemes, the current scheme can be applied without intensive training to deliver satisfying annotation results. Experience may improve rating, suggested by the higher reliability and lower rating uncertainty of the experienced coder. The easy-to-apply scheme can be broadly utilised in research and clinical settings for efficient initial annotation of infant vocalisations. The generated datasets further suit further in-depth manual and computational analyses, paving the path to examine emerging verbal functions – deep acoustic phenotyping – of infants of different developmental and clinical phenotypes.

## Einleitung

Das Verständnis der Ontogenese menschlichen Verhaltens beginnend mit der fötalen Entwicklung, in welcher die Grundlagen für viele komplexe Fähigkeiten und Kapazitäten später im Leben gebildet werden, hat sich in den letzten Forschungsjahrzehnten dramatisch erweitert (z.B. Einspieler et al., 2021a, b). Mit zu den faszinierendsten Fragen der Entwicklungsforschung zählt, ob, wann und wie sich interindividuelle Unterschiede manifestieren und welche frühen Fähigkeiten auf die weitere Entwicklung projizieren. Eine entwicklungsdynamische Sichtweise, die frühe Funktionen als Vorläufer und Voraussetzungen für spätere ansieht, geht davon aus, dass auch frühe Abweichungen oder Beeinträchtigungen späteren suboptimalen Merkmalen oder unerwünschten Entwicklungsverläufen vorausgehen, selbst wenn die Kernsymptomatik bestimmter Störungen erst später auftritt (z.B. Autismus-Spektrum-Störung). Diese auch als Neurokonstruktivismus bezeichnete Sichtweise (z.B. Karmiloff-Smith, 2009; Mareschal, 2011; Westermann et al., 2011) ist eng verknüpft mit dem Bestreben frühe funktionelle Marker für Neurodiversität, d.h. Prädiktoren für Entwicklungsverläufe, zu finden (z.B. Karmiloff-Smith, 2009; Marschik et al., 2017). Hinweise auf mögliche Marker leiten sich einerseits aus dem direkten Vergleich mit Gleichaltrigen (*Peers*), einem verlangsamten oder abweichenden Entwicklungsverlauf oder uneinheitlichen Entwicklungsprofilen her (z.B. Diskrepanzen in einzelnen Entwicklungsdomänen – Motorik, Kognition, Sozialentwicklung, Sprechen-Sprache-Kommunikation). Der sprachlichen Domäne kommt dabei oft eine besondere Bedeutung zu, denn obwohl die frühkindliche Sprachentwicklung noch nicht restlos verstanden oder vollständig erforscht ist, sind häufig Auffälligkeiten in diesem Entwicklungsbereich die ersten Anzeichen einer abweichenden Entwicklung oder Anzeichen dafür, eine Entwicklungsdiagnostik zu initiieren.

Die frühe Vokalisationsentwicklung wurde seit langem als Abfolge von Kompetenzschritten dargestellt und bereits vor mehr als 40 Jahren in sogenannten Stufenmodellen beschrieben (ursprünglich als *Stage Models* benannt; für neuere Übersichten siehe Kent, 2022; Oller, 2000; Marschik et al., 2022). Ganz zu Beginn produzieren Neugeborene Schreie, vegetative Laute (z.B. Schluckauf, Niesen) und kurze, unmodulierte, stimmhafte Vokalisationen (*Quasi Resonant Sounds*). Ab einem Alter von einem Monat werden Vokalisationen zunehmend länger und es kommen erste artikulatorische Elemente während der Phonation hinzu (Gurren, *Cooing*). Mit etwa drei Monaten erreichen Kinder die "Expansionsphase", in der stimmliche und artikulatorische Möglichkeiten exploriert werden. Dies äußert sich im Auftreten von verschiedensten vokal- und konsonantähnlichen Vokalisationen, die isoliert oder auch in ersten Sequenzen, dem sogenannten "marginalen Babbeln", produziert werden. Ab dem Alter von fünf Monaten tritt schließlich das "kanonische Babbeln" auf, welches durch schnelle Formantübergänge in Konsonant-Vokal-Silben bzw. die Produktion von Silbenketten gekennzeichnet ist. Die Untersuchungslage zur frühkindlichen Vokalisationsentwicklung in den ersten 5 Lebensmonaten und damit vor Beginn des kanonischen Babbelns ist vergleichsweise gering. Eine in den letzten Jahren immer mehr in den Vordergrund drängende Möglichkeit zur Erforschung frühkindlicher Vokalisationen ist der Einsatz computergestützter akustischer Verfahren und sogenannter *Machine Learning Ansätze* (Marschik et al., 2022). Bis dato gibt es einige akustische Beschreibungen früher Lautproduktion in den ersten Lebensmonaten (cf. Schreien: z.B. Wermke et al., 2007; Protophone: z.B. Oller et al., 2019; kategorienübergreifend: Buder et al., 2013), die jedoch nicht oder nur in Einzelfällen als Klassifizierungsgrundlage (Zuordnung zu Vokalisationsklassen) dienten. Unterschiede zwischen Jungen und Mädchen in dieser frühen Phase waren primär quantitativ, interessanterweise produzieren Jungen mehr Lautäußerungen als Mädchen, qualitativ zeigen sich jedoch keine Unterschiede (Sung et al., 2013; Gratier et al. 2015; Oller et al., 2020).

Ziel dieser Arbeit ist es, an Hand eines Datensatzes einer neurotypischen Säuglings-Kohorte ein Kodierschema zu entwickeln, um in weiterer Folge detaillierte menschliche und computergestützte Analysen von frühkindlichen Vokalisationen im physiologischen und klinischen Kontext zu ermöglichen. Dieses "Grund-Kodierschema" soll (i) einfach durchführbar (i.e. intuitiv und ohne spezifisches zeitintensives Training anwendbar) und reliabel sein; und (ii) die daraus resultierenden Datensätze sollen universell einsetzbar und weiter verarbeitbar sein (d.h. vergleichbare Korpora zur künftigen linguistischen Transkription sowie automatisierten Analyse generieren). Um die Nützlichkeit und das Einsatzpotenzial des neuen Kodierschemas zu zeigen, präsentieren wir einfache computergestützte Anwendungen, mit dem Ziel in der Literatur berichtete Befunde wie z.B. zu Geschlechtsunterschieden in der Vokalisationsentwicklung zu replizieren.

## Methode

Weiterführende Details zu den Einschlusskriterien, der Durchführung (Vorsegmentierung, Familiarisierung mit dem Kodierschema und Konsolidierung der Ratings, Kodierungsprozess, akustische Analyse etc.) entnehmen Sie bitte dem Elektronischen Supplement (ES).

### *Stichprobe, Studiendesign und Audiomaterial*

Im Rahmen der prospektiven Längsschnittstudie "Early Human Development: the 3-Month Transformation" (Marschik et al., 2017; Reich et al., 2021; Krieber-Tomantschger et al., 2022) mit Fokus auf die typische neuromotorische, visuelle und verbale Entwicklung wurden für die folgende Studie Daten von 45 Säuglingen (23 weiblich) ausgewählt. Insgesamt wurde jedes Neugeborene sieben Mal untersucht (am 28., 42., 56., 70., 84., 98. und 112. Tag nach errechnetem Geburtstermin) und das Spontanverhalten mittels Audio-Videoaufzeichnungen aufgenommen. Die Gesamtdauer verwertbarer Audiodaten betrug 1.501 Minuten und beinhaltete 9.305 analysierbare Segmente (Details siehe ES).

*Kodierschema*

Aufbauend auf gängigen Kodierschemata frühkindlicher Vokalisationen (z.B. Buder et al., 2013; Nathani et al., 2006; Oller, 2000; Oller et al., 2013) stellen wir hier ein Konzept eines "Grund-Kodierschemas" vor, das als Grundlage für weitere Detailannotationen und computergestützte Analysen dienen soll. Im Gegensatz zu den oben genannten Schemata, die darauf abzielen kategoriale Zuordnungen für kindliche Lautäußerungen zu treffen (z.B. Gurren, Babbeln oder beispielsweise eine Unterscheidung in *Squeals* und *Growls*; Oller, 2000; Oller et al., 2013), bezieht sich das hier vorgestellte Schema auf phonetische Grundeigenschaften von Lautäußerungen, weshalb man es auch als "Grund-Kodierung oder Grund-Annotation" (Layer 1) bezeichnen kann. Um eine weitere Detailanalyse (menschlich oder computergestützt; Layer 2) zu ermöglichen, haben wir alle präsegmentierten „nicht stillen Segmente" (siehe ES) fünf Layer 1 Hauptklassen zugeordnet: (1) **Stimmhafte Lautäußerung**: teilweise oder insgesamt mit Stimmbeteiligung produzierte und gegebenenfalls im Ansatzrohr artikulatorisch modulierte Laute; (2) **Stimmlose Lautäußerung**: Vokalisationen ohne Stimmbeteiligung. (3) **Definiertes Signal** (*Fixed Signals*; Oller, 2000): positive (z.B. Lachen oder Kichern) oder negative (z.B. Weinen oder Quengeln) Lautäußerungen sowie vegetative Laute (z.B. Niesen). (4) **Non-Target**: Laute und Geräusche, die nicht von dem Kind stammen oder nicht vokalisiert sind (z.B. Hintergrundgeräusche; Geräusche, die durch kindliche Bewegungen entstehen – nicht relevant zur Analyse kindlicher Vokalisationen). (5) **Nicht zuordenbar**: Lautäußerungen, die vom Kind stammen, aber nach dreimaligem Anhören (siehe auch Kodierungsprozess, ES) von Kodierern nicht eindeutig zu den Klassen (1)-(3) zugeordnet werden können.

*Analyse akustischer Eigenschaften frühkindlicher Vokalisationen (Replikation bestehender Erkenntnisse)*

Von den Segmenten, die den einzelnen Klassen des Grund-Kodierschemas von beiden Kodiererinnen einheitlich zugewiesen wurden (siehe Ergebnisse), untersuchten wir zunächst einfache akustische Parameter und deren Verteilung. Hierzu analysierten wir die durchschnittliche Dauer von Lautäußerungen, getrennt für Jungen und Mädchen, wobei nur Lautäußerungen der Klassen (1) Stimmhafte Lautäußerung und (2) Stimmlose Lautäußerung betrachtet wurden. Zudem bestimmten wir u.a. die Grundfrequenz, um Geschlechtsunterschiede in den frühen Vokalisationen zu ermitteln.

# Ergebnisse

*Intrarater Reliabilität*

Kodiererin 1 erreichte für die Doppelannotation (siehe ES) der 200 randomisiert ausgewählten Segmente ein Kappa von *k* = 0,82 (95% Konfidenzintervall [0,75-0,89]) auf alle fünf Klassen bezogen. Kodiererin 2 erzielte auf dem deutlich reduzierten Set von 92/200 Segmenten (siehe oben) eine Übereinstimmung von *k* = 0,68 (95% Konfidenzintervall [0,56-0,80]).

*Interrater Reliabilität*

Die Übereinstimmung zwischen den beiden Kodiererinnen wurde für 9.305 Segmente, die von beiden annotiert wurden, berechnet. Details sind in der Konfusionsmatrix, in der die Verteilung der von den zwei Kodiererinnen vergebenen Klassen abgebildet ist, dargestellt (Abbildung 1). Der Cohen's Kappa für alle fünf Klassen betrug 0,70 (95% Konfidenzintervall [0,69-0,71]). Die Gesamtdiskrepanz, d.h. die Nicht-Übereinstimmung, betrug 1.635 Segmente (17,6% von 9.305; Segmente in allen Zellen, die in Abbildung 1 nicht auf der Diagonale von links oben nach rechts unten sind). Davon wurden 1.140 (12,3% von 9.305) von einer der Kodiererinnen der Klasse (5) zugeordnet. Das bedeutet, während eine Kodiererin sich sicher war und eine Klasse gewählt hat, wählte die andere Kodiererin die Klasse (5) aus Unsicherheit. Entschieden sich beide Kodiererinnen für eine der anderen Klassen (1)-(4), waren lediglich 495 Segmente (5,3% von 9.305) unterschiedlich zugewiesen. In Summe wurden 1.602 Segmente (17,2% von 9.305) von mindestens einer Kodiererin der Klasse (5) "Nicht zuordenbar" zugewiesen. Das bedeutet, diese Vokalisationen stammen vom Kind, können aber nicht sicher einer der Klassen (1)-(3) zugewiesen werden. Von Kodiererin 1 wurden 728 Segmente (7,8%) und von Kodiererin 2 1.140 Segmente (12,3%) der Klasse (5) „Nicht zuordenbar" zugewiesen. Davon waren 266 (2,9%) überlappend. Wenn Klasse (5) herausgenommen wurde (s.a. Abbildung 1), lag die Interrater Reliabilität für die verbleibenden vier Klassen von 7.703 Segmenten (82,78%) bei *k* = 0,92 (95% Konfidenzintervall [0,91-0,93]).

Aus von beiden Kodiererinnen übereinstimmend annotierten 7.670 Segmenten (82,4%; Summe der Segmente in den Zellen aus Abbildung 1 von links oben diagonal nach rechts unten) wurden 2.565 Segmente (27,6%) der Klasse (1) "Stimmhafte Lautäußerung", 10 Segmente (0,1%) der Klasse (2) "Stimmlose Lautäußerung", 135 Segmente (1,5%) der Klasse (3) "Definiertes Signal", 4.694 Segmente (50,4%) der Klasse (4) "Non-Target" und 266 Segmente (2,9%) der Klasse (5) "Nicht zuordenbar" zugewiesen (s.a. Tabelle 1). Die Dauer aller Segmente der Klassen (1) und (2) betrug 987,16 Sekunden. In der weiteren Analyse inkludierten wir nur die von beiden Kodiererinnen übereinstimmend annotierten Segmente.

|  | Stimmhafte Lautäußerung | Stimmlose Lautäußerung | Definiertes Signal | Non-Target | Nicht Zuordenbar |
|---|---|---|---|---|---|
| Stimmhafte Lautäußerung | **2565** 27.6% | 2 0.02% | 54 0.6% | 131 1.4% | 403 4.3% |
| Stimmlose Lautäußerung | 2 0.02% | 10 0.1% | 0 0% | 3 0.03% | 13 0.1% |
| Definiertes Signal | 33 0.4% | 0 0% | 135 1.5% | 14 0.2% | 54 0.6% |
| Non-Target | 41 0.4% | 15 0.2% | 4 0.04% | 4694 50.4% | 404 4.3% |
| Nicht Zuordenbar | 139 1.5% | 12 0.1% | 12 0.1% | 299 3.2% | 266 2.9% |

(Kodierer 1 = Zeilen, Kodierer 2 = Spalten)

*Abbildung 1: Konfusionsmatrix der Kodierung/Annotation.*

*Tabelle 1: Klassen mit der zugehörigen Anzahl und durchschnittlicher Dauer einzelner Segmente von den 7.670 übereinstimmend annotierten Segmenten.*

| Klasse | Anzahl | Totale Dauer der Äußerungen (Sek.) | Durchschnittliche Dauer (+/- Standardabweichung) und [Median] in Sekunden |
|---|---|---|---|
| (1) Stimmhafte Lautäußerung | 2565 | 983,97 | 0,38 (+/- 0,32) [*0,29*] |
| (2) Stimmlose Lautäußerung | 10 | 3,19 | 0,32 (+/- 0,17) [*0,26*] |
| (3) Definiertes Signal | 135 | 88,12 | 0,65 (+/- 0,52) [*0,46*] |
| (4) Non-Target | 4694 | 1244,64 | 0,27 (+/- 1,16) [*0,15*] |
| (5) Nicht zuordenbar | 266 | 184,84 | 0,69 (+/- 1,91) [*0,21*] |

**Geschlechtsunterschiede bei frühkindlichen Lautäußerungen**

Wir untersuchten die Parameter Dauer und Grundfrequenz der Lautäußerungen getrennt für Jungen und Mädchen. Zuerst quantifizierten wir die Länge der Audiosegmente getrennt für Jungen und Mädchen. Hierzu nahmen wir alle Segmente der Klassen (1) "Stimmhafte Lautäußerung" und (2) „Stimmlose Lautäußerung" (insgesamt 2575 Segmente). Lautäußerungen dauerten bei Jungen im Durchschnitt 0,39 Sekunden (Standardabweichung = 0,32). Die kürzeste Lautäußerung lag bei 0,07 Sekunden und die längste bei 2,73 Sekunden. Lautäußerungen von Mädchen dauerten im Durchschnitt 0,38 Sekunden (Standardabweichung = 0,31). Die kürzeste Lautäußerung war 0,07 Sekunden lang und die längste 2,90 Sekunden. Die Lautierungsdauer ist nicht signifikant unterschiedlich zwischen beiden Geschlechtern (Two-Sample t-Test, $p$ = 0,5603; siehe Abbildung 2, links).

Die durchschnittliche Grundfrequenz der Lautäußerungen bei Mädchen betrug 371.10 Hz (Standardabweichung = 64.09), wobei die geringste Grundfrequenz bei 67.64 Hz und die höchste bei 511.33 Hz lag. Die Lautäußerungen von Jungen zeigten im Schnitt eine Grundfrequenz von 366.53 Hz (Standardabweichung = 65.92), wobei die geringste Grundfrequenz 67.33 Hz und die höchste 518.74 Hz betrug. Die Grundfrequenzen sind somit ebenso nicht signifikant unterschiedlich zwischen beiden Geschlechtern ($p$ = 0,1845; siehe Abbildung 2, rechts).

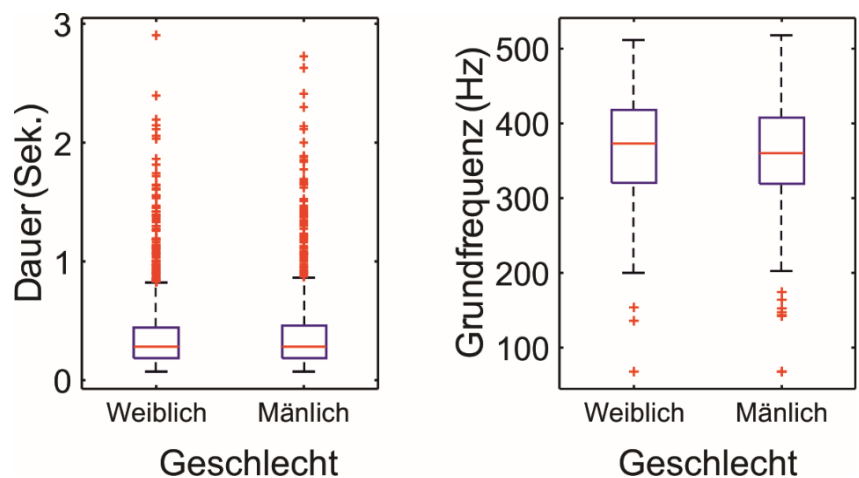

*Abbildung 2: (Links) Boxplot der Dauer der Lautäußerungen nach Geschlecht (Anzahl der Segmente w: 1214; m: 1361). (Rechts) Boxplots der Grundfrequenz der Lautäußerungen nach Geschlecht (Anzahl der Segmente w: 648; m: 803). Für die Auswertung der Grundfrequenz wurden alle Segmente exkludiert für die keine Grundfrequenz approximiert werden konnte. Die Whiskers zeigen den Interquartilsabstand (IQA) mal 1.5 an. Schlüssel: ● - Outlier > 1.5 IQA.*

## Diskussion

Ziel der Studie war es, anhand des Datensatzes von 45 neurotypischen Säuglingen ein einfach durchführbares und reliables Kodierschema frühkindlicher Vokalisationen zu entwickeln und zu prüfen. Dieses soll weiterführende verhaltensanalytische und computergestützte Analysen von frühkindlichen Vokalisationen, Deep Acoustic Phenotyping, im physiologischen und klinischen Kontext ermöglichen. Per Algorithmus automatisiert segmentierte Audiosignale wurden durch zwei Kodiererinnen den fünf Klassen: (1) **Stimmhafte Lautäußerung**, (2) **Stimmlose Lautäußerung**, (3) **Definiertes Signal**, (4) **Non-Target** und (5) **Nicht zuordenbar** zugewiesen. Um die Verwertbarkeit des Datensatzes nach Anwendung des "Grund-Kodierungsprozesses" (Layer 1 Annotation) beispielhaft zu zeigen, wurden die von beiden Kodiererinnen identisch gelabelten Segmente mit computergestützten Methoden phonetisch analysiert. Die Erkenntnisse und Ergebnisse lassen sich wie folgt interpretieren: In der Familiarisierungsphase benötigten beide Kodiererinnen lediglich 2x100 Audiosegmente um sich mit dem Konzept des Kodierschemas und der technischen Umsetzung vertraut zu machen. Die Konsolidierungsphase mit drei sich nicht überlappenden randomisierten Datensätzen zu je 100 Segmenten ergab schon bei der ersten Durchführung für jeden der Datensätze eine hohe Übereinstimmung in der Klassenzuordnung (1)-(5) zwischen den Kodiererinnen. Es war somit kein intensives Training notwendig, um sich mit dem neuen Kodierschema vertraut zu machen und es erwies sich für beide Kodiererinnen als intuitiv und unkompliziert in der Anwendung trotz unterschiedlicher Kodierungserfahrungen. Diese Erkenntnis ist bedeutend, da sie im Gegensatz zu anderen Kodierschemata steht, für welche häufig intensive Trainings vor der Kodierung der Untersuchungsdaten durchgeführt werden müssen (z.B. Nathani et al., 2006; Oller et al., 2019). Als Unterschied zu diesen Schemata ist allerdings hervorzuheben, dass unser vorgeschlagenes System eine „Grund-Kodierung" (Layer 1, s.u.) darstellt.

Für die 9.305 von beiden Kodiererinnen annotierten Segmente erhielten wir eine vergleichbare Interrater Reliabilität wie andere Studien, die ähnliche Schemata verwendeten (z.B. Zappella et al., 2015; Legerstee et al., 1992). Wir generierten eine zusätzliche Klasse (5) "Nicht zuordenbar", in die alle vom Kind stammenden Vokalisationen, die nicht eindeutig den Klassen (1)-(3) zugewiesen werden können, eingingen. Anders als Klasse (4), mit der alle nicht vom Kind stammenden Segmente gelabelt werden, enthalten die Klassen (1)-(3) und Klasse (5) Vokalisationen, die eindeutig dem Kind zugewiesen werden können. Die Klasse (5) ermöglicht dabei perzeptive Unsicherheit auszudrücken, ohne gezwungen zu sein ein Segment einer Klasse/Kategorie zuzuordnen. Vom Gesamtdatensatz der 9.305 Segmente wurden 1.635 (17,6%) unterschiedlich von den Kodiererinnen kategorisiert. Der größte Anteil dieser Nicht-Übereinstimmung (70%) resultiert aus der "Unsicherheit" einer der beiden Kodiererinnen (d.h. eine der beiden labelte diese Segmente als "Nicht zuordenbar"; siehe Abbildung 1). Wenn die Klasse (5) aus den Berechnungen exkludiert wurde, lag der

Übereinstimmungswert für Zuordnungen zu den Klassen (1)-(4) bei 0,92. Dies bedeutet, dass die Interrater-Reliabilität für das vorgeschlagene Kodierschema bei sicherer Zuordnung der Kodiererinnen nahezu perfekt war. „Nicht zuordenbare" Vokalisationen haben unter Umständen akustische Merkmale (reduzierte Lautstärke, Überlagerung des Hintergrundrauschens, extreme Kürze), welche Kodierer*innen vor Herausforderungen stellen und eindeutige Zuordnungen erschweren. Diese Segmente, die von zumindest einer Kodiererin als "Nicht zuordenbar" (5) klassifiziert wurden, sollen in künftigen Studien genauer analysiert werden, um zu erheben, ob die darin enthaltenen kindlichen Lautäußerungen gegebenenfalls artikulatorisch oder stimmlich anders produziert wurden als jene der Klassen (1)-(3), oder ob sie von externalen Faktoren verzerrt wurden, was die auditive Perzeption beeinflusst.

Es ist wichtig zu betonen, dass Kodiererin 1 mit mehr Kodierungserfahrung wesentlich weniger Segmente der Klasse (5) zuwies (7,8%) als Kodiererin 2 ohne Kodierungserfahrung (12,3%). Die Intrarater Reliabilität zeigte auch eine höhere Stabilität bei der erfahrenen ($k$ = 0,82) gegenüber der unerfahrenen ($k$ = 0,68) Kodiererin. Obwohl also das vorgeschlagene "Grund-Kodierschema" intuitiv, einfach und ohne viel Training anwendbar ist, zeigt sich in der Anwendung – wie für jedes andere Kodierschema auch erwartbar – eine gewisse Beeinflussung durch kodiererspezifische Eigenschaften (z.B. Vorerfahrung, generelles Selbstbewusstsein, Balance zwischen Kodierungsgeschwindigkeit und -genauigkeit). Im Gegensatz zu anderen Studien haben wir jedoch nicht einfach Segmente für die weitere Analyse (inklusive Reliabilität) exkludiert, die nicht sicher einer Klasse zugewiesen werden konnten (z.B. D'Odorico et al., 2011), oder haben die Kodiererinnen "gezwungen", eine Entscheidung zu treffen. Klasse (5) ermöglichte den Kodiererinnen, explizit ihre Unsicherheit zu benennen und dadurch den Grund der Nicht-Übereinstimmung transparent und differenziert im Ergebnis zu präsentieren. Ob dieser Kodiervorgang, der Unsicherheit zulässt, im Gegensatz zu anderen Vorgängen ohne eine solche Kategorie (also nur Klassen (1)-(4) in diesem Schema), die Kodierstile und damit die Resultate beeinflusst, muss in weiterführenden Studien geprüft werden.

Wie bereits erwähnt stellt unser Schema einen ersten Bearbeitungsschritt dar, auf den aufbauend weitere Klassifikationen getroffen werden können, während viele andere Schemata verschiedene Vokalisationstypen unterscheiden. Die beiden Klassen (1) "Stimmhafte Lautäußerung" und (2) "Stimmlose Lautäußerung" sind dabei als Sprachvorläufer bzw. Protophone zu verstehen, während Vokalisationen der Klasse (3) "Definiertes Signal" als "*Non-Speech-Like*" (Oller, 2000) gelten.

Hinsichtlich des Ziels dieser Studie lässt sich zusammenfassend das Fazit ziehen, dass die hier vorgestellte "Grund-Kodierung/Annotation" weitgehend intuitiv und ohne spezifisches zeitintensives Training mit guten Intrarater- und Interrater Übereinstimmungen anwendbar ist. Die Unterteilung in "Grund-Kodierung" (Layer 1) und Detailanalyse (Layer 2) stellt dabei eine Neuheit und je nach Studienziel eine potenzielle Erleichterung gegenüber bisherigen Kodierschemata dar. Zum einen ermöglicht die "Grund-Kodierung" als erster von zwei Bearbeitungsschritten unabhängig von dem Erfahrungsschatz der kodierenden Person eine sehr effiziente erste Klassifikation hinsichtlich der weiter zu verarbeitenden Daten. Der weitere Schritt, die Detailanalyse, welche entweder (teil-)automatisiert oder durch erfahrene Kodierer und Kodiererinnen erfolgen kann, wird dadurch um viele nicht zu analysierende Laute (Non-Target) bereinigt. Datensätze, die mit diesem "Grund-Kodierschema" vorklassifiziert sind, können somit von Kodierer*innen hinsichtlich weiterer Vokalisationstypen oder auf Signalebene (z.B. Vergleich bestimmter Gruppen: Geschlecht, Alter, typische vs. atypische Entwicklung; Parameter für Deep Acoustic Phnotyping) weiterbearbeitet werden.

Aufbauend auf der "Grund-Kodierung" wurden die beiden Klassen (1) "Stimmhafte Lautäußerung" und (2) "Stimmlose Lautäußerung" in der vorliegenden Studie weiter analysiert. Beispielhaft wurden die durchschnittliche Dauer von Lautäußerungen sowie die Grundfrequenz, getrennt für Jungen und Mädchen bestimmt. Die Ergebnisse replizieren vorangegangene Studien, die zeigten, dass sich frühkindliche Vokalisationen von Mädchen und Jungen hinsichtlich dieser Parameter nicht unterschieden (z.B. Sung et al., 2013; Gratier et al. 2015).

Im Gegensatz zur Einordnung des kindlichen Vokalisationsrepertoires anhand von Stufenmodellen (Oller, 2000; Nathani et al., 2006; Kent, 2022; Marschik et al., 2022) steht bei der Anwendung des vorgeschlagenen Schemas nicht unbedingt die Suche nach dem Erreichen von Meilensteinen im Vordergrund, sondern die weiterführende Analyse der Qualität kindlicher Vokalisationen, ggf. eine dimensionale oder kategoriale Klassifikation, oder die Bestimmung altersspezifischer akustischer Parameter.

Wurde bis dato die frühkindliche Sprachentwicklung überwiegend verhaltensanalytisch/ perzeptiv untersucht und beschrieben, gibt es zunehmend mehr Studien, die akustische Parameter (mit)erheben (z.B. Warlaumont et al., 2010; Pokorny et al., 2018; Oller et al. 2019; Buder et al 2013; Überblick in Marschik et al., 2022). So verwendeten z.B. Warlaumont et al. (2010) Spektra als Eingabe in ein neuronales Netz, um innerhalb der Klasse von Protophonen (d.h. *Vocant, Squeal, Growl*; Oller et al., 2013) Unterschiede zu finden. Neben diesen Beschreibungen typischer akustischer Merkmale konnten wir z.B. in unseren vorangegangenen Studien Parameter wie "*Timbre*" oder "*Loudness*" als potenzielle Marker, die zur Unterscheidung zwischen sich typisch entwickelnden Kindern und Kindern mit Neuroentwicklungsstörung beitragen könnten, beschreiben (z.B. Pokorny et al., 2018; 2022). In der hier präsentierten Studie mit neurotypischen Kindern untersuchen wir die Faktoren Grundfrequenz und Dauer, da diese in zahlreichen Studien in Bezug zu altersspezifischen Vokalisationen erhoben wurden (z.B. Sohner & Mitchell, 1991; Lynch et al., 1995; Brisson et al., 2014; Marschik et al., 2022). Diese zwei Faktoren stellen lediglich einen ersten Schritt in unserem Analysevorhaben dar, akustische Parameter in ganz frühen Lautäußerungen im Detail zu beschreiben (Deep Acoustic Phenotyping). Wie in unseren Vorarbeiten beschrieben (z.B. Marschik et al., 2017; 2022; Pokorny et al., 2018; 2022) planen wir zukünftig weitere Parameter auch für diese frühen Vokalisationen in den ersten Lebensmonaten zu erheben (z.B. Timbre, Jitter, MFCCs; c.f. über prädefinierte Parametersets wie z.B. eGeMAPS oder ComParE; Schuller et al., 2013, Eyben et al., 2015). Diese Parametersets haben wir bis dato unter anderem in der Analyse von Vokalisationen in der Phase des kanonischen Babbelns verwendet (z.B. Pokorny et al., 2017; 2018; 2022).

Bisherige Studien zur prä-kanonischen Entwicklung bei Kindern mit Neuroentwicklungs-störungen sind selten und die Ergebnisse zu Verhaltensunterschieden zwischen sich typisch und atypisch entwickelnden Kindern wenig eindeutig (z.B. Lang et al., 2019; 2021; Roche et al., 2018; Yankowitz et al., 2019; Marschik et al., 2022). Beobachtete Atypikalitäten in den ersten 5 Lebensmonaten waren z.B. das Nichterreichen des Entwicklungsmeilensteins "Gurren" bei Kindern mit Autismus Spektrum Störung (Maestro et al., 2002; Zappella et al., 2015); typische Vokalisationen im wechselweisen Auftreten mit atypischen, z.B. inspiratorischen, Vokalisationen bei Kindern mit Rett Syndrom (Marschik et al., 2009; 2012), abweichende Intonationskonturen und weniger stimmliche Reaktionen in interaktiven Situationen bei Kindern mit Autismus Spektrum Störung (Brisson et al., 2014). Unsere Übersichtsarbeit (Marschik et al., 2022) zeigt, dass die Ergebnisse bisheriger Studien nicht nur rar, sondern oft auch abhängig von den angewandten Methoden ausfallen und aufgrund der meist geringen Stichprobengrößen in ihrer Aussage beschränkt bleiben. Einem der Gründe für solch geringe Stichprobengrößen in bisherigen Studien, dem sehr hohen zeitlichen Aufwand bei ausschließlich manueller Segmentierung und Kodierung/Annotation, kann durch das hier vorgeschlagene Vorgehen künftig entgegengewirkt werden. Ein Ausbau der Effizienz im Segmentierungs- und Kodierungsprozess wird die Anwendung im Forschungssetting sowie im klinischen Kontext positiv beeinflussen und den Aufbau großer Datenkorpora ermöglichen. Eine Kombination aus automatisierten und manuellen Methoden kann den Rahmen bieten, die detaillierte Untersuchung der frühkindlichen Vokalisationsentwicklung voranzutreiben, Atypikalitäten zu erkennen sowie Stillstände oder Rückschritte in der Entwicklungskaskade genauer zu ergründen. Die weitere Suche in der frühen verbalen Entwicklung nach funktionellen Markern für Neurodiversität und somit Prädiktoren für Unterschiede in individuellen Entwicklungsverläufen sollte Ziel intensiver zukünftiger Forschung sein.

## Danksagung

## Referenzliste

**Supplement**

In den meisten Studien zur frühkindlichen Sprachentwicklung werden überwiegend Beobachtungsverfahren angewendet, bei denen "menschliche Kodierer" das Datenmaterial (Audio- oder Videoaufnahmen) zunächst segmentieren und anschließend annotieren. D.h. im ersten Schritt werden Einheiten, die für die weitere Analyse verwendbar sind, selektiert und in einem zweiten Schritt werden die kindlichen Lautäußerungen auditiv klassifiziert und vordefinierten Kategorien zugeordnet. Die Segmentierung kindlicher Lautäußerungen erfolgt vielfach über Atemzyklen (Lynch et al., 1995). Dies bedeutet, wenn eine Lautäußerung von einer Inspirationsphase unterbrochen wird, dann wird dies als Segmentgrenze angegeben auf die eine neue Lautäußerung folgt. Eine andere Möglichkeit Vokalisationen zu "trennen" sind sogenannte Pausenkriterien: überschreitet eine "Ruhephase" (Stille, keine Lautgebung) zwischen zwei Lautäußerungen einen vorher definierten Schwellenwert (z.B. Pausen länger als 300ms; Oller et al., 2010; Pausen länger als 2s; Nathani et al., 2006), stellt dies eine Segmentgrenze und somit eine Trennung zwischen Vokalisationen dar. Grundsätzliches Problem der manuellen Segmentierung und Annotation ist der sehr hohe zeitliche Aufwand und eine gewisse Fehlerhaftigkeit, die eine breite Anwendung im klinischen und auch Forschungssetting erschwert (z.B. Oller et al., 2010). Selbst in Studien zur typischen frühkindlichen Sprachentwicklung wurden und werden aufgrund dieser Schwierigkeit häufig nur kleinere Probandengruppen untersucht, weshalb Ergebnisse oft schwer generalisierbar sind.

Oller und Kollegen untersuchten diese frühe Phase und entwickelten ein Kodierungssystem, welches kindliche Lautäußerungen in (a) Schreien/Weinen, (b) Lachen und (c) Protophone unterteilt. Letztere, die auch als Vorläufer von Sprache definiert sind, wurden weiter in "Vocants", "Squeals" und "Growls" gegliedert (Jhang & Oller, 2017; Oller et al., 2013). Die entscheidenden Beobachtungen für Vokalisationen dieser Klasse war, dass sie bereits innerhalb der ersten Lebensmonate in unterschiedlichen Kontexten und dabei variabel verwendet werden, also eine funktionale Flexibilität vorliegt. Außerdem belegten diese Studien, dass bereits in den ersten Lebensmonaten nicht Schreien sondern Protophone die vorherrschende Vokalisationsart sind (Jhang & Oller, 2017; Oller et al., 2013).

**Entwicklung eines neuen Kodierschemas**

In erster Linie ging es in der vorliegenden Arbeit darum, die Grundlagen für ein Kodierschema zu entwickeln, das es ermöglicht, Vokalisationen in den ersten 4 Lebensmonaten (korrigiertes Alter) zu identifizieren und zu klassifizieren (z.B. Lautäußerungen mit und ohne Stimmbeteiligung oder "definierte Signale"; *Fixed Signals*; wie z.B. Lachen, Weinen etc.; Details siehe Methode), etc. Diese grundlegenden Kategorien bilden die Basis für jede weitergehende Analyse und genaue Untersuchung der vorsprachlichen Vokalisationen.

**Methode**

*Stichprobe, Studiendesign und Audiomaterial*

Im Rahmen unserer prospektiven Längsschnittstudie "Early Human Development: the 3-Month Transformation" (Krieber-Tomantschger et al., 2022; Marschik et al., 2017; Reich et al., 2021) mit Fokus auf die typische neuromotorische, visuelle und verbale Entwicklung wurden zwischen 2015 und 2017 einundfünfzig Säuglinge (26 weiblich) untersucht. Einschlusskriterien waren eine unauffällige Schwangerschaft, eine komplikationslose termingerechte (>37 Wochen Gestationsalter) Geburt sowie Neonatalperiode. Nicht eingeschlossen wurden Kinder, die ein älteres Geschwisterkind mit Neuroentwicklungsstörung oder einer Kinder-Jugendpsychiatrischen Auffälligkeit hatten, sowie Kinder von Eltern mit psychiatrischen, genetischen oder metabolischen Vorerkrankungen. Die Studie wurde von der Ethikkommission der Medizinischen Universität Graz (27-476ex14/15) sowie von der Ethikkommission der Universitätsmedizin Göttingen (IDENTIFIED) positiv votiert und von allen Eltern liegt eine schriftliche Einverständniserklärung zur Studienteilnahme und Publikation der Ergebnisse vor. Von den 51 zunächst eingeschlossenen Kindern wurde eines aufgrund einer mit 3 Jahren diagnostizierten genetischen Erkrankung ausgeschlossen, bei fünf weiteren gibt es fehlende Datenpunkte, die für die Auswertungen dieser Studie notwendig sind. In die hier vorliegende Teilerhebung der Längsschnittstudie flossen somit Erhebungen von insgesamt 45 (23 weiblich) sich typisch entwickelnden monolingual deutschsprachig aufwachsenden Kindern ein.

Ab dem Alter von 4 Wochen nach dem errechneten Geburtstermin wurden die Spontanmotorik, die spontanen Vokalisationen (nicht interaktiv) und die visuelle Aufmerksamkeit der Probanden in vierzehntägigen Intervallen erhoben. Insgesamt wurde jedes Neugeborene sieben Mal untersucht (am 28., 42., 56., 70., 84., 98. und 112. Tag nach errechnetem Geburtstermin, jeweils mit max. +/- 2 Tagen zulässiger Abweichung; für Details siehe Marschik et al., 2017). Die Säuglinge wurden im Verhaltenszustand "wach und aktiv" (*Active Wakefulness*) (ursprüngliches Konzept zurückgehend auf Prechtl, 1974) zur Erhebung des frühkindlichen Spontanverhaltens in Rückenlage für einige Minuten in ein Gitterbett gelegt und mittels eines multimodalen multisensorischen Messaufbaus aufgenommen. Details zum Messaufbau finden Sie ebenso in Marschik et al. (2017) (generelles Studiendesign) und Reich et al. (2021) (Spontanmotorik). Für die vorliegende Studie wurden die jeweiligen Audiosequenzen aus der multimodalen Messung zur Analyse des Spontanverhaltens im Detail analysiert (vom Video wurden reine Audio-Segmente extrahiert, die über Kopfhörer präsentiert wurden). Die Gesamtdauer aller verfügbaren Videoaufnahmen der 45 eingeschlossenen Kinder betrug 2.859 Minuten, durchschnittlich 9 Minuten und 39 Sekunden je Aufnahme. In diesem Zeitfenster waren zum Teil Sequenzen, in denen die Mutter/der Vater das Kind in das Gitterbett legt oder wieder aufnimmt enthalten, wie auch ggf. Phasen, in denen das Kind quengelt und von einem Elternteil beruhigt wurde. Da Ziel dieser Studie die Analyse der Spontanvokalisationen der Neugeborenen war, wurden von einer Forschungsassistentin alle Videos im Detail angesehen und Sequenzen in denen das Kind generell (über einen längeren Zeitraum) quengelig, schlafend oder "nicht allein" war, d.h. ein Elternteil mit dem Kind interagierte, herausgefiltert.

### *Durchführung*

#### *Automatische Vorsegmentierung*

Das gesamte auf Spontanverhalten "bereinigte" Audiomaterial (siehe oben) wurde mittels Schwellenwert-Algorithmus weiterbearbeitet und alle "Stillen Segmente" (d.h. Signale < 25dB) wurden markiert (Python-Script). Diese Schwelle wurde bewusst gesetzt, um möglichst alle kindlichen Vokalisationen herauszufiltern und möglichst wenige zu verwerfen. Dadurch blieb neben analysierbaren Segmenten zunächst eine größere Anzahl von "Signalen" im Datensatz enthalten (zur Weiterverarbeitung siehe Annotation). "Signale" sind einerseits Vokalisationen des Kindes aber auch Geräusche über dem Schwellwert von 25 dB (z.B. Geräusche, die durch frühkindliche Bewegungen – Spontanmotorik, Einspieler et al., 2008 – entstehen; Umgebungsgeräusche inklusive Hintergrundrauschen; Sprache im Hintergrund, wie z.B. Fragen der Eltern an die Testleitung). Des Weiteren definierten wir "Stille Segmente" von >100ms als "Pause". Ein "Nicht stilles Segment" ist somit eine "Einheit" zwischen zwei aufeinanderfolgenden Pausen. Der Verarbeitungsschritt „Automatische Vorsegmentierung" ergab 9.305 "Nicht stille Segmente". Diese Segmente wurden für den nächsten Verarbeitungsschritt, die Kodierung (Layer 1 Annotation), exportiert.

#### *Kodierungsprozess*

Zwei menschliche Kodiererinnen waren in den Kodierungsprozess, für den eine eigens mit Python

programmierte Software zur Verfügung stand (Basisevaluation, Pre-Processing, mittels Noldus Observer XT), involviert. Kodiererin 1 hat eine logopädische Grundausbildung, war jahrelang als Sprachtherapeutin tätig und hat weitreichende Expertise in der Analyse frühkindlicher Vokalisationen. Kodiererin 2 ist eine Psychologieabsolventin ohne jegliche Vorerfahrung im Bereich frühkindliche Sprachentwicklung.

*Familiarisierung:* Vertraut machen mit dem Kodierschema: Beide Kodiererinnen wurden über die fünf Layer 1 Hauptklassen des Kodierschemas: (1) Stimmhafte und (2) Stimmlose Lautäußerung; (3) Definiertes Signal; (4) Non-Target; (5) Nicht zuordenbar (siehe Manuskript) instruiert. Ein randomisiert ausgewähltes Set von 100 Segmenten aus dem Gesamtdatensatz (N = 9.305) wurde jeder Kodiererin zur Verfügung gestellt, um sich mit der Aufgabe, den fünf Klassen und der technischen Umsetzung (Kodierungsprogramm) vertraut zu machen. Jedes Audiosegment durfte 3-mal abgespielt werden bevor die Kodiererinnen die jeweilige Zuteilung zu den Klassen durch Drücken einer definierten Taste festlegten. Etwaige Fragen wurden gesammelt und im Team diskutiert. Weitere Sets zu je 100 randomisiert gewählten Segmenten konnten zur Familiarisierung genutzt werden, bis die Kodiererin die entsprechende Sicherheit über das Konzept und den Ablauf (manuelle Umsetzung) erlangte. Beide Kodiererinnen benötigten zwei Familiarisierungssets um diese Sicherheit zu erlangen (2 x 100 Segmente) und mit der Konsolidierung zu beginnen.

*Konsolidierung*: Drei nicht überlappende (*Mutual Exclusive Samples*) und randomisiert gewählte Sets zu je 100 Segmenten wurden den Kodiererinnen präsentiert. Ziel dieses Schritts war es, in drei aufeinanderfolgenden Sets eine Übereinstimmung zwischen beiden Kodiererinnen (*Interrater Agreement*) von Cohen's Kappa ($k$) = 0,80 oder höher zu erzielen. Jede Kodiererin annotierte die Sets unabhängig und für jedes Set wurde das Kappa zwischen den Kodiererinnen in den fünf Klassen berechnet. Beiden Kodiererinnen wurde jeweils nach fertiger Annotation eines Sets das Ergebnis mitgeteilt, bevor das nächste Set annotiert wurde. Für alle diese drei Durchläufe lag die Übereinstimmung zwischen den Kodiererinnen für fünf Klassen über einem Kappa von 0,80. Wäre dies nicht der Fall gewesen, hätten neue Sets generiert werden und so lange konsolidiert werden müssen, bis das Kriterium von drei aufeinanderfolgenden Übereinstimmungswerten über 0,80 erreicht ist. Nachdem dies jedoch bereits mit den ersten drei Sets erreicht war, wurde die Konsolidierungsphase abgeschlossen.

*Formale Annotation zur Bestimmung der Ground-Truth (dt. Grundwahrheit)*: Aus den 9.305 Segmenten (inklusive der 500 Segmente aus der Familiarisierungs- und Konsolidierungsphase) wurden

200 Segmente zufällig ausgewählt und als Duplikate zum Gesamtsample dazu gefügt (9.305 + 200), um später die Intrarater Reliabilität zu überprüfen. Diese 9.505 Segmente wurden erneut vollständig randomisiert. Geplant war es, diese 200 Segmente von jeder Kodiererin je zweimal annotieren zu lassen. Jede Kodiererin konnte selbst über den zeitlichen Ablauf des Kodierungsprozesses bestimmen. Es fand kein Austausch zwischen den Kodiererinnen statt. Der annotierte Datensatz von Kodiererin 2 umfasst 9.397 Segmente (davon sind 92 der 200 zweifach annotierten Segmente). Kodiererin 1 annotierte den vollständigen Datensatz (9.505 Segmente inklusive der 200 doppelt vorkommenden Segmente).

**Akustische Analyse**

Die Grundfrequenz wurde mittels Python Bibliothek *librosa* berechnet. Bis auf die Abtastfrequenz des Eingabesignals, welche 44,1 kHz betrug und die Fensterlänge für die wir 93ms verwendeten, wurden alle anderen Eingabeparameter wie von der Bibliothek vorgeschlagen übernommen (d.h. *hop_length = frame length / 4, win_length= frame length / 2, fmin = 65.41 Hz, fmax = 2093.00 Hz*). Wir nutzten den Algorithmus "Probabilistic YIN" (Mauch & Dixon, 2014), was darin resultierte, dass für 43.6% der Segmente aus den Klassen (1) und (2) keine Grundfrequenz approximiert werden konnte, und somit in dieser Analyse nicht verwendbar waren. Dies liegt daran, dass das Signal keine eindeutige Grundfrequenz aufwies, da entweder die Dauer des Segments zu kurz oder die Signal-to-Noise Proportion zu hoch war.

**Referenzliste (Zusatz)**